*Article*

# Thick Cloud Removal of Remote Sensing Images Using Temporal Smoothness and Sparsity-Regularized Tensor Optimization

**Chenxi Duan [1], Jun Pan [1,\*], Rui Li [2]**

[1] State Key Laboratory of Information Engineering in Surveying, Mapping and Remote Sensing, Wuhan University, 129 Luoyu Road, Wuhan 430079, China; chenxiduan@whu.edu.cn(C.D.); panjun1215@whu.edu.cn (J.P.)

[2] School of Remote Sensing and Information Engineering, Wuhan University, Wuhan 430079, China; lironui@whu.edu.cn

\* Correspondence: panjun1215@whu.edu.cn

**Abstract:** In remote sensing images, the presence of thick cloud accompanying cloud shadow is a high probability event, which can affect the quality of subsequent processing and limit the scenarios of application. Hence, removing the thick cloud and cloud shadow as well as recovering the cloud-contaminated pixels is indispensable to make good use of remote sensing images. In this paper, a novel thick cloud removal method for remote sensing images based on temporal smoothness and sparsity-regularized tensor optimization (TSSTO) is proposed. The basic idea of TSSTO is that the thick cloud and cloud shadow are not only sparse but also smooth along the horizontal and vertical direction in images while the clean images are smooth along the temporal direction between images. Therefore, the sparsity norm is used to boost the sparsity of the cloud and cloud shadow element, and unidirectional total variation (UTV) regularizers are applied to ensure the unidirectional smoothness. This paper utilizes alternation direction method of multipliers (ADMM) to solve the presented model and generate the cloud and cloud shadow element as well as the clean element. The cloud and cloud shadow element is purified to get the cloud area and cloud shadow area. Then, the clean area of the original cloud-contaminated images is replaced to the corresponding area of the clean element. Finally, the reference image is selected to reconstruct details of the cloud area and cloud shadow area using the information cloning method. A series of experiments are conducted both on simulated and real cloud-contaminated images from different sensors and with different resolutions, and the results demonstrate the potential of the proposed TSSTO method for removing cloud and cloud shadow from both qualitative and quantitative viewpoints.

**Keywords:** cloud removal; group sparsity; unidirectional total variation (UTV); tensor optimization;

## 1. Introduction

Remote sensing images have been widely used in many research and application fields, such as classification [1,2], object detection [3], and urban geographical mapping [4]. However, remote sensing images are usually affected by cloud cover. It is estimated that the global average cloud cover is 66% by the International Satellite Cloud Climatology Project (ISCCP) [5]. Thick clouds usually cover up the earth's surface information, so the information on cloud-contaminated pixels are always inaccurate or missing, which notably affects usability and limits further analysis of remote sensing images. Since thick cloud cover is inevitable for optical remote sensing images and influences the subsequent processing, using reasonable methods to remove the cloud-contaminated area is imperative.

Thick clouds usually lead to land cover information completely lost, so the basic idea of removing thick clouds is to recover the cloud-contaminated pixels using available information. To this end, many methods for thick cloud removal have been proposed in the last few decades. According to the source of information exploited to reconstruct the cloud-cover area, the existing methods can be classified into spatial information-based methods, spectral information-based methods, and temporal information-based methods, which are described in more detail in the following.

◼ Spatial information-based methods

The spatial information-based methods are dedicated to recovering the cloud-contaminated pixels by making full use of the information from the cloud-free area in the repairing image. Missing pixel interpolation is the elementary method of these methods [6,7], which is only suitable for dealing with tiny missing areas but cannot deal with the large area absence. To overcome this limitation, some researchers utilized new techniques for cloud image restoration [8-10]. Criminisi inpainting [8] fulfilled the missing area within the target image and reconstructed pixels in an appropriate order patch by patch. Bandelet-based inpainting utilized the spectro-geometrical information in the cloud-free parts of the cloud-contaminated image [9]. Maximum a posteriori (MAP)-based method got cloud-free image using apriori knowledge and gradient descent optimization [10]. Recently, to better get the desired image, Cheng et al. [11] and Zeng et al. [12] proposed methods based on similar pixel locations and global optimization to get seamless image. Besides, tensor completion [13] such as FaLRTC and HaLRTC also can be generalized to cloud removal and cloud shadow removal in multitemporal remotely sensed images. However, spatial information-based methods can only get visually reasonable cloud-free image when the cloud-contaminated area is small, once the missing area becomes larger, the results of these methods are not practicable for subsequent quantitative analysis and applications [14].

◼ Spectral information-based methods

The main idea of the spectral information-based methods can be described as follows. Certain bands that with stronger penetrating force than the cloud-contaminated bands can avoid bad effects from cloud to some extent. This means the information contained in clean bands can be exploited to fill the cloud-contaminated area. Specifically, Zhang et al. [15] represented a haze optimized transformation (HOT) method which corrected the visible bands radiometrically and removed the cloud in Landsat images. What is more, to make use of MODIS band 7 to remove the cloud in MODIS bands 6, many researchers modeled the relationship between the two bands [16-18]. Besides, Li et al. [19] made the utmost use of shortwave infrared imagery to defog for optical imagery, and Malek et al. [20] recently harnessed autoencoder networks to reconstruct the cloud area of the multispectral image. Although spectral information-based methods can consider more information related to the cloud area, they are invalid when all the bands are contaminated by thick clouds.

◼ Temporal information-based methods

Temporal information-based methods use the pixels in reference images corresponding to the cloud-contaminated pixels in the target images. As enough reference information is taken into consideration, high effectiveness is the main advantage of temporal information-based methods. When the cloud-contaminated area is wider, the superiority of this kind of method is more obvious. One of the most commonly used methods is managing to build a linear relationship between cloud-contaminated pixels in the target image and corresponding pixels in the reference image [21-22]. The main problem that needs to be addressed is the obvious stitching traces. Researchers presented many ways out to get seamless results. For example, Zeng et al. [23] and Li et al. [5] proposed different residual correction methods to guarantee seamless reconstruction. Since the spectral differences and land cover changes between multitemporal images increased the difficulty of cloud removal, many scholars took different measures to keep the spectral consistency as much as possible. Cheng et al. [14] constructed cloud area using similar pixels in the reference image, which was constrained by a spatio-temporal Markov random field global function. Patch-based information reconstruction was also proposed to minimize the effect of the land cover changes [25,26]. Besides, Zhang et al. [27] introduced Deep Learning (DL) framework into the missing data reconstruction field to extract multiscale features and for information recovery. Another group of popular methods among temporal information-based methods is matrix factorization. Regarding the cloud is sparse and the cloud-free element possesses the characteristic of low-rank, Wen et al. [28] and Zhang et al. [29] introduced the robust principal component analysis (RPCA) into the cloud removal task. By applying discrepant RPCA, they proposed two-pass RPCA (TRPCA) and group-sparsity constrained RPCA (GRPCA) to detect and remove the cloud. Another type of sparse representation method, which is becoming frequently-used, is dictionary learning [30-33]. For instance, based on patch matching-based multitemporal group sparse representation, the local temporal correlations and the nonlocal spatial correlations were used to reconstruct the missing data [32]. Similar to sparse representation, Li et al. [34] proposed a nonnegative matrix factorization-based method but the coefficients of which were not always sparse. Furthermore, many researchers expanded this series of methods from two-dimensional matrix to high-dimensional tensor, even combining them with total variation (TV). Ng et al. [35] and Cheng et al. [36] designed an adaptive weighted tensor completion method to recover the cloud-contaminated pixels. Ji et al. [37] presented a nonlocal tensor completion method to restore the missing area. Chen et al. [38] proposed TVLRSDC, a low-rank sparsity decomposition method with total variation regularized, which could remove the cloud and cloud shadow simultaneously. Generally, the temporal information-based methods can obtain satisfactory results, as long as the spectral difference and the land cover change is not too large between the target image and reference image.

Different from the above-mentioned method, this paper proposes a novel tensor optimization model based on temporal smoothness and sparse representation (TSSTO) for thick cloud and cloud shadow removal in remote sensing images. The proposed TSSTO method is mainly based on the discriminatively intrinsic features that the cloud and cloud shadow are usually sparse and smooth along the horizontal and vertical direction in images while the clean images are smooth along the temporal direction between images. The main contributions of the presented paper could be listed as follows:

- This paper ensures temporal smoothness by unidirectional total variation regularizer, which avoids expensive SVD computation. Considering the gradient along the temporal direction between images in the clean element is smaller than that in the cloud and cloud shadow element, unidirectional total variation (UTV) regularizer along the temporal direction is exploited to constrain the temporal smoothness of the clean element.
- Group sparsity is used to enhance the sparsity of the cloud and cloud shadow, and two unidirectional total variation regularizers along the horizontal direction and vertical direction are designed to deal with the large cloud and cloud shadow. This enables TSSTO to better remove both the large and small cloud-contaminated area.

The rest of this paper is organized as follows. In Section 2, the proposed TSSTO method and implementation details are introduced. In Section 3, a series of experiments are conducted, and the results of TSSTO are compared with those of contrast methods. Besides, the performance of the proposed method is analyzed and the superiority of the proposed method is discussed. The conclusions of this paper and the direction of future work are reported in Section 4.

## 2. Methodology

In this paper, for most images that cover the non-tropical regions and that do not have sudden significant surface changes between images, the thick cloud and cloud shadow element is sparse and smooth along the horizontal direction and vertical direction while the clean images are smooth along the temporal direction between images. Based on this prior knowledge, the cloud and cloud shadow element as well as the clean element is obtained by tensor optimization. The cloud and cloud shadow element is constrained by the group sparsity and two UTV regularizers along the horizontal direction and vertical direction. The temporal smoothness of the clean element is guaranteed by the UTV regularizer along the temporal direction. As the tensor optimization changes the pixel values in the original clean area, the pixels in the clean area in the original images are replaced to the clean element.

To recover the clearer texture, details are constructed by information cloning to get cloud removal results.

As shown in Figure 1, the proposed TSSTO method contains three steps: Firstly, the images can be arranged to a tensor according to the acquisition time. By executing tensor optimization based on the proposed model, the suspected cloud and cloud shadow element as well as the clean element can be acquired. Since the clean element is obtained from the original cloud-contaminated image and the cloud with cloud shadow has been removed, the clean element can be seen as the preliminary recovery results. Secondly, thresholds are applied to the cloud and cloud shadow element to get the cloud area and cloud shadow area. Then, according to the cloud area and cloud shadow area, the clean area in the original images is substituted for the corresponding area in the preliminary recovery result. And this step together with the tensor optimization step is called TOS. Thirdly, to recover texture information more completely, the details of the cloud area and cloud shadow area are recovered by information cloning. It is worth noting that the cloud and cloud shadow will be removed only when the clean complementary information is available in the multitemporal images. The input of the proposed TSSTO method is a sequence of images that cover the same area, and the cloud masks or cloud shadow masks could be provided or not.

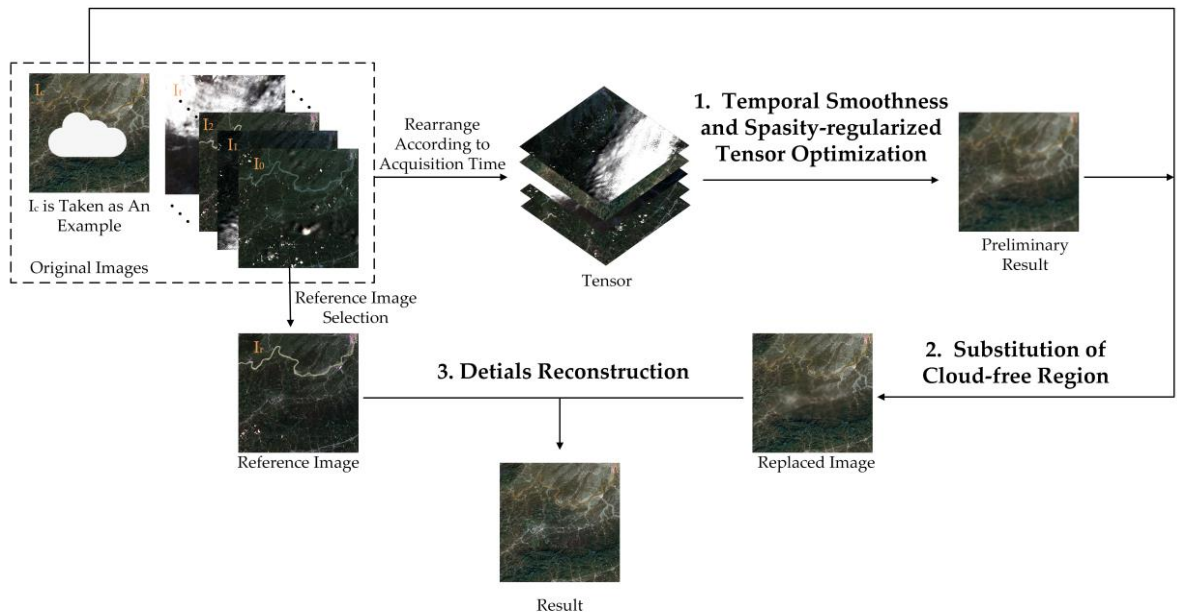

**Figure 1.** Flowchart of the proposed TSSTO method.

*2.1. Tensor Optimization Based on Temporal Smoothness and Sparsity*

Generally, if the cloud is not heavy, the cloud and cloud shadow element is supposed to be sparser than the clean images in the cloud-contaminated images. Therefore, group sparsity is used to promote the sparsity of the cloud and cloud shadow element.

To better explain the UTV regularizers, Figure 2 shows the original cloud-contaminated images, clean element, as well as the cloud and cloud shadow element. And Figure 3-Figure 5 show the histogram of the absolute values of the gradients in different directions.

As shown in Figure 3-Figure 4, the horizontal gradient and vertical gradient of the cloud and cloud shadow element are generally smaller than those of the clean images. Hence, it is reasonable to constrain the horizontal and vertical gradient of the cloud and cloud shadow element to be smaller. In the temporal direction, as shown in Figure 5, the gradient of the clean images are usually small, and the temporal gradient of the cloud and cloud shadow element tends to be very large or very small. The gradient of the cloud and cloud shadow area in the cloud and cloud shadow element tends to be very large, while the gradient of the clean area in the cloud and cloud shadow element tends to be very small. This means the most important information of the cloud and cloud shadow is in the part of the large gradient. By constraining the temporal gradient of the clean element to be small, the cloud and cloud shadow element can be obtained, and the general spectrum of the cloud area and cloud shadow area can be recovered in the clean element.

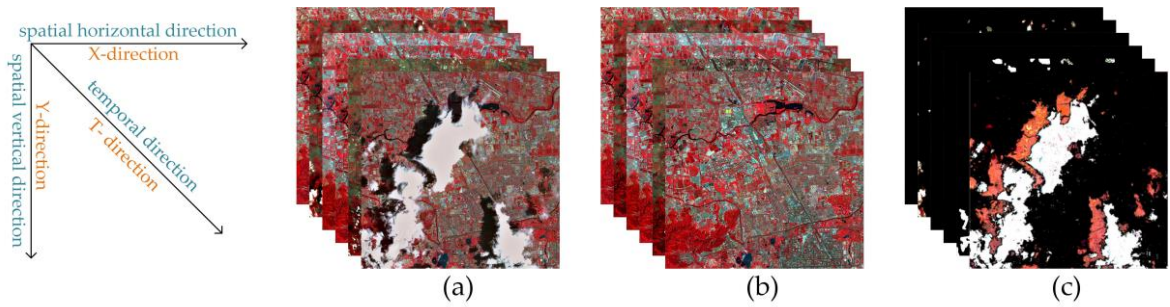

**Figure 2.** The images used to draw histograms (a) original cloud-contaminated images; (b) clean element; (c) cloud with cloud shadow element.

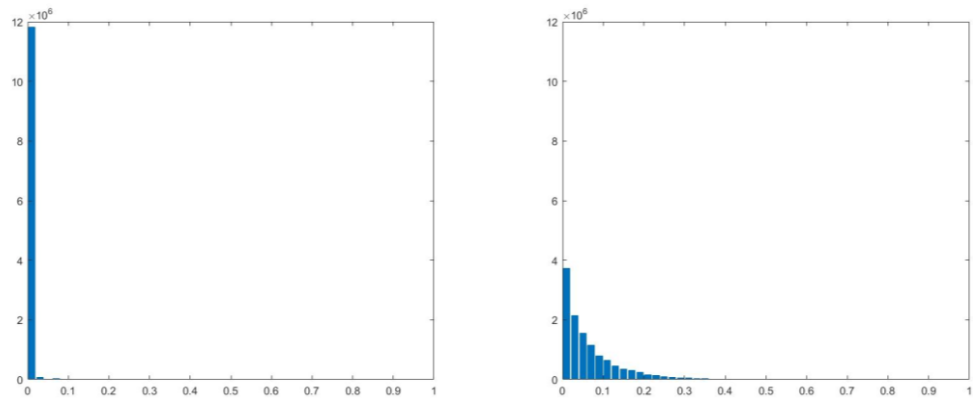

**Figure 3.** The histogram of the absolute values of the horizontal gradient of (a) cloud with cloud shadow element and (b) clean element.

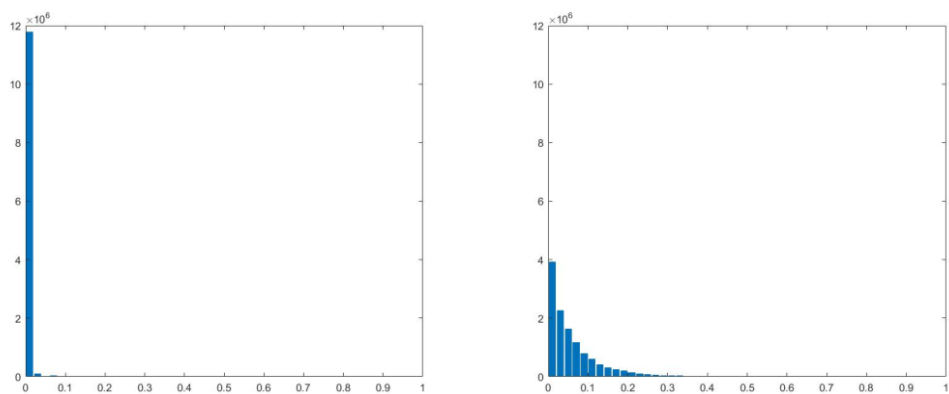

**Figure 4.** The histogram of the absolute values of the vertical gradient of (a) cloud with cloud shadow element and (b) clean element.

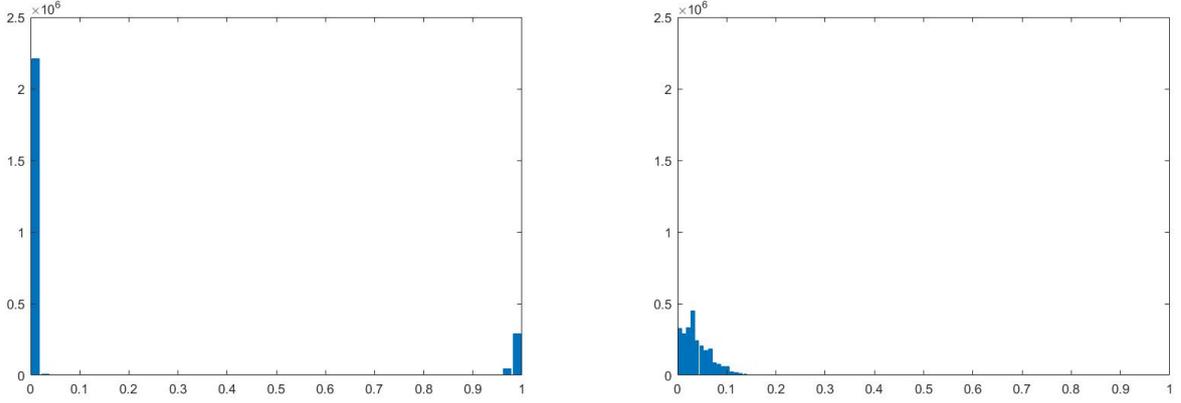

**Figure 5.** The histogram of the absolute values of the temporal gradient of (a) cloud with cloud shadow element and (b) clean element.

2.1.1. Temporal Smoothness and Sparsity-Regularized Model

Mathematically, a single remote sensing image can be represented as $I \in \mathbb{R}^{m \times n \times b}$ where $m$ and $n$ denote the numbers of rows and columns, and $b$ denotes the number of spectral bands, respectively. Given a series of remote sensing images, they can be arranged into a tensor, $\mathcal{D} \in \mathbb{R}^{m \times n \times b \times t}$, where $t$ denotes the numbers of images, and thus the images can be processed band by band.

Supposing that $\mathcal{D}, \mathcal{C}$, and $\mathcal{B} \in \mathbb{R}^{m \times n \times t}$ are used to represent the original cloud-contaminated images, the cloud and cloud shadow element, and the clean element, respectively. The cloud removal model can be formulated as

$$\arg \min_{\mathcal{B},\mathcal{C}} \quad \lambda_1 \|\nabla_x \mathcal{C}\|_1 + \lambda_2 \|\nabla_y \mathcal{C}\|_1 + \lambda_3 \|\nabla_z \mathcal{B}\|_1 + \lambda_4 \|\mathcal{C}\|_{2,1} \tag{1}$$

$$\text{s.t.} \quad \mathcal{D} = \mathcal{B} + \mathcal{C}, \quad \mathcal{B} \geq 0.$$

where $\nabla_x$, $\nabla_y$, and $\nabla_z$ denote the derivative operators of horizontal dimension, vertical dimension, and temporal dimension, respectively. $\lambda_1$, $\lambda_2$, $\lambda_3$, and $\lambda_4$ are the parameters that adjust the weight of different information sources.

The terms $\|\nabla_x \mathcal{C}\|_1$ and $\|\nabla_y \mathcal{C}\|_1$ are unidirectional total variation (UTV) regularizers to enhance the smoothness of the cloud and cloud shadow element along the horizontal direction and vertical direction. The derivatives of the cloud and cloud shadow element along the vertical direction and horizontal direction are sparser than the derivatives of clean images, which is shown in Figure 3 and Figure 4.

The term $\|\nabla_z \mathcal{B}\|_1$ is the temporal smoothness term, which is applied because of a similar tendency for the clean images as time goes on. Generally, for the cloud and cloud shadow element, there is little correlation in the temporal domain between images as a result of the randomness of the thick cloud. Consequently, the cloud and cloud shadow element usually have large gradients in the temporal domain between images. Contrary to the cloud and cloud shadow element, the clean images own small temporal gradients because of just changing a little according to a certain trend over time.

The term $\|\mathcal{C}\|_{2,1}$ represents the group sparsity of the cloud and cloud shadow element. When the cloud is not heavy, the cloud and cloud shadow can be seen as a sparse term approximately. Normally, the cloud and cloud shadow element is sparser than the clean element for a useful image. Hence, $\|\mathcal{C}\|_{2,1}$, a $\ell_{2,1}$ norm, is used to promote the group sparsity of the cloud and cloud shadow element. For a matrix $X$, $\|X\|_{2,1}$ is defined as

$$\|X\|_{2,1} = \sum_{i=1}^{s} \|x_{g_i}\|_2 \tag{2}$$

where $g_i$ denotes the $i$th group of the variable $X$. And $s$ is the number of groups. For a tensor $\mathcal{C}$, $\|\mathcal{C}\|_{2,1}$ is defined as

$$\|\mathcal{C}\|_{2,1} = \|\text{unfold}_1(\mathcal{C})\|_{2,1} \tag{3}$$

2.1.2. Optimization of the Proposed Model

To obtain $\mathcal{B}$ and $\mathcal{C}$ in Equation (1) effectively, the constrained equation is rewritten to fit the alternating direction method of multipliers (ADMM) framework [40]:

$$\arg\min_{\mathcal{A},\mathcal{C},\mathcal{H},\mathcal{V},\mathcal{T}} \lambda_1\|\mathcal{H}\|_1 + \lambda_2\|\mathcal{V}\|_1 + \lambda_3\|\mathcal{T}\|_1 + \lambda_4\|\mathcal{A}\|_{2,1}$$

$$\text{s.t.} \quad \mathcal{A} = \mathcal{C}$$
$$\mathcal{H} = \nabla_x \mathcal{C}$$
$$\mathcal{V} = \nabla_y \mathcal{C} \tag{4}$$
$$\mathcal{T} = \nabla_z (\mathcal{D} - \mathcal{C})$$
$$\mathcal{D} \geq \mathcal{C},$$

Let $\mathcal{Y} = [\mathcal{Y}_1, \mathcal{Y}_2, \mathcal{Y}_3, \mathcal{Y}_4]$ be the Lagrange multiplier and $\mu$ be a positive penalty scalar. The augmented Lagrange function [39] can be defined as follows.

$$\begin{aligned} f_\mu(\mathcal{A},\mathcal{C},\mathcal{H},\mathcal{V},\mathcal{T},\mathcal{Y}) = & \lambda_1\|\mathcal{H}\|_1 + \lambda_2\|\mathcal{V}\|_1 + \lambda_3\|\mathcal{T}\|_1 + \lambda_4\|\mathcal{A}\|_{2,1} \\ & + \langle \mathcal{Y}_1, \mathcal{A} - \mathcal{C} \rangle + \frac{\mu}{2}\|\mathcal{A} - \mathcal{C}\|_F^2 \\ & + \langle \mathcal{Y}_2, \mathcal{H} - \nabla_x \mathcal{C} \rangle + \frac{\mu}{2}\|\mathcal{H} - \nabla_x \mathcal{C}\|_F^2 \\ & + \langle \mathcal{Y}_3, \mathcal{V} - \nabla_y \mathcal{C} \rangle + \frac{\mu}{2}\|\mathcal{V} - \nabla_y \mathcal{C}\|_F^2 \\ & + \langle \mathcal{Y}_4, \mathcal{T} - \nabla_z (\mathcal{D} - \mathcal{C}) \rangle + \frac{\mu}{2}\|\mathcal{T} - \nabla_z (\mathcal{D} - \mathcal{C})\|_F^2 \end{aligned} \tag{5}$$

And the Equation (5) is equivalent to

$$\begin{aligned} f_\mu(\mathcal{A},\mathcal{C},\mathcal{H},\mathcal{V},\mathcal{T},\mathcal{Y}) = & \lambda_1\|\mathcal{H}\|_1 + \lambda_2\|\mathcal{V}\|_1 + \lambda_3\|\mathcal{T}\|_1 + \lambda_4\|\mathcal{A}\|_{2,1} \\ & + \frac{\mu}{2}\left\|\mathcal{A} - \mathcal{C} + \frac{\mathcal{Y}_1}{\mu}\right\|_F^2 \\ & + \frac{\mu}{2}\left\|\mathcal{H} - \nabla_x \mathcal{C} + \frac{\mathcal{Y}_2}{\mu}\right\|_F^2 \\ & + \frac{\mu}{2}\left\|\mathcal{V} - \nabla_y \mathcal{C} + \frac{\mathcal{Y}_3}{\mu}\right\|_F^2 \\ & + \frac{\mu}{2}\left\|\mathcal{T} - \nabla_z (\mathcal{D} - \mathcal{C}) + \frac{\mathcal{Y}_4}{\mu}\right\|_F^2 \end{aligned} \tag{6}$$

Separating the variables $[\mathcal{A},\mathcal{C},\mathcal{H},\mathcal{V},\mathcal{T}]$ into two groups which are $[\mathcal{A},\mathcal{H},\mathcal{V},\mathcal{T}]$ and $[\mathcal{C}]$, the ADMM framework can be used to figure out the equation (4) by optimizing the following five sub-problems.

$$\begin{aligned} \mathcal{A}^{k+1} &= \arg\min_{\mathcal{A}} f_\mu(\mathcal{A}, \mathcal{C}^k, \mathcal{H}^k, \mathcal{V}^k, \mathcal{T}^k, \mathcal{Y}^k) \\ \mathcal{C}^{k+1} &= \arg\min_{\mathcal{C}} f_\mu(\mathcal{A}^k, \mathcal{C}, \mathcal{H}^k, \mathcal{V}^k, \mathcal{T}^k, \mathcal{Y}^k) \\ \mathcal{H}^{k+1} &= \arg\min_{\mathcal{H}} f_\mu(\mathcal{A}^k, \mathcal{C}^k, \mathcal{H}, \mathcal{V}^k, \mathcal{T}^k, \mathcal{Y}^k) \\ \mathcal{V}^{k+1} &= \arg\min_{\mathcal{V}} f_\mu(\mathcal{A}^k, \mathcal{C}^k, \mathcal{H}^k, \mathcal{V}, \mathcal{T}^k, \mathcal{Y}^k) \\ \mathcal{T}^{k+1} &= \arg\min_{\mathcal{T}} f_\mu(\mathcal{A}^k, \mathcal{C}^k, \mathcal{H}^k, \mathcal{V}^k, \mathcal{T}, \mathcal{Y}^k) \end{aligned} \tag{7}$$

and the Lagrange multiplier $\mathcal{Y}$ can be updated as

$$\begin{aligned}
\mathcal{Y}_1^{k+1} &= \mathcal{Y}_1^k + \mu(\mathcal{A}^k - \mathcal{C}^k) \\
\mathcal{Y}_2^{k+1} &= \mathcal{Y}_2^{k+1} + \mu(\mathcal{H}^k - \nabla_x \mathcal{C}^k) \\
\mathcal{Y}_3^{k+1} &= \mathcal{Y}_3^{k+1} + \mu(\mathcal{V}^k - \nabla_y \mathcal{C}^k) \\
\mathcal{Y}_4^{k+1} &= \mathcal{Y}_4^{k+1} + \mu(\mathcal{T}^k - \nabla_z(\mathcal{D} - \mathcal{C}^k))
\end{aligned} \tag{8}$$

Following, the approach of updating variables in these five sub-problems will be introduced.

**update $\mathcal{A}$**:

$$\mathcal{A}^{k+1} = \arg\min_{\mathcal{A}} \lambda_4 \|\mathcal{A}\|_{2,1} + \frac{\mu}{2}\left\|\mathcal{A} - \mathcal{C}^k + \frac{\mathcal{Y}_1}{\mu}\right\|_F^2 \tag{9}$$

The above problem is equivalent to

$$\mathcal{A}_{gi}^{k+1} = \arg\min_{\mathcal{A}} \sum_{g_i}\left(\min_{\mathcal{A}} \lambda_4 \|\mathcal{A}_{gi}\|_{2,1} + \frac{\mu}{2}\|\mathcal{A}_{gi} - \mathcal{R}_{gi}^k\|_F^2\right)$$

$$\mathcal{R}_{g_i} := \mathcal{C} - \frac{(\mathcal{Y}_1)_{gi}}{\mu} \tag{10}$$

which can be solved by the soft thresholding formula:

$$\mathcal{A}_{gi}^{k+1} = \max\left(\|\mathcal{R}_{gi}^k\|_2 - \frac{\lambda_4}{\mu}, 0\right)\frac{\mathcal{R}_{gi}^k}{\|\mathcal{R}_{gi}^k\|_2} \tag{11}$$

where $\mathcal{A}_{gi}^{k+1}$ denotes the *i*th group of the variable $\mathcal{A}^{k+1}$.

**update $\mathcal{C}$**:

$$\begin{aligned}
\mathcal{C}^{k+1} = \arg\min_{\mathcal{C}} &\frac{\mu}{2}\left\|\mathcal{A}^{k+1} - \mathcal{C} + \frac{\mathcal{Y}_1}{\mu}\right\|_F^2 + \frac{\mu}{2}\left\|\mathcal{H}^k - \nabla_x \mathcal{C} + \frac{\mathcal{Y}_2}{2}\right\|_F^2 \\
&+ \frac{\mu}{2}\left\|\mathcal{V}^k - \nabla_y \mathcal{C} + \frac{\mathcal{Y}_3}{\mu}\right\|_F^2 + \frac{\mu}{2}\left\|\mathcal{T}^k - \nabla_z(\mathcal{D} - \mathcal{C}) + \frac{\mathcal{Y}_4}{\mu}\right\|_F^2
\end{aligned} \tag{12}$$

is a least square problem, and the normal equation can be written as

$$\begin{aligned}
&\left(\mu I + \mu \nabla_x^T \nabla_x + \mu \nabla_y^T \nabla_y + \mu \nabla_z^T \nabla_z\right)\mathcal{C}^{k+1} \\
&= \mu \mathcal{A}^{k+1} + \mathcal{Y}_1 + \nabla_x^T(\mu \mathcal{H}^k + \mathcal{Y}_2) + \nabla_y^T(\mu \mathcal{V}^k + \mathcal{Y}_3) \\
&+ \nabla_z^T(\mu \nabla_z \mathcal{D} - \mu \mathcal{T}^k - \mathcal{Y}_4)
\end{aligned} \tag{13}$$

which can be calculated by a closed-form solution as follows

$$\mathcal{C}^{k+1} = \mathcal{F}^{-1}\left(\frac{\mathcal{L}_1}{\mathcal{L}_2}\right) \tag{14}$$

where $\mathcal{L}_1 = \mu \mathcal{A}^{k+1} + \mathcal{Y}_1 + \nabla_x^T(\mu \mathcal{H}^k + \mathcal{Y}_2) + \nabla_y^T(\mu \mathcal{V}^k + \mathcal{Y}_3) + \nabla_z^T(\mu \nabla_z \mathcal{D} - \mu \mathcal{T}^k - \mathcal{Y}_4)$

$\mathcal{L}_2 = \mu I + \mu \nabla_x^T \nabla_x + \mu \nabla_y^T \nabla_y + \mu \nabla_z^T \nabla_z$

**update $\mathcal{H}, \mathcal{V}, \mathcal{T}$**:

$$\mathcal{H}^{k+1} = \arg\min_{\mathcal{H}} \lambda_1 \|\mathcal{H}\|_1 + \frac{\mu}{2}\left\|\mathcal{H} - \nabla_x \mathcal{C}^{k+1} + \frac{\mathcal{Y}_1}{\mu}\right\|_F^2 \tag{15}$$

$$\mathcal{V}^{k+1} = \arg\min_{\mathcal{H}} \lambda_2 \|\mathcal{V}\|_1 + \frac{\mu}{2}\left\|\mathcal{V} - \nabla_y \mathcal{C}^{k+1} + \frac{\mathcal{Y}_2}{\mu}\right\|_F^2 \tag{16}$$

$$\mathcal{T}^{k+1} = \arg\min_{\mathcal{H}} \lambda_3 \|\mathcal{T}\|_1 + \frac{\mu}{2}\left\|\mathcal{T} - \nabla_z(\mathcal{D} - \mathcal{C}^{k+1}) + \frac{\mathcal{Y}_3}{\mu}\right\|_F^2 \tag{17}$$

as in [41], component-wise soft thresholding is an exact updating solution for the above three problems

$$\mathcal{H}^{k+1} = \mathcal{S}_{\frac{\lambda_1}{\mu}}\left(\nabla_x \mathcal{C}^{k+1} - \frac{\mathcal{Y}_1}{\mu}\right) \tag{18}$$

$$\mathcal{V}^{k+1} = \mathcal{S}_{\frac{\lambda_2}{\mu}}\left(\nabla_y \mathcal{C}^{k+1} - \frac{\mathcal{Y}_2}{\mu}\right) \tag{19}$$

$$\mathcal{T}^{k+1} = \mathcal{S}_{\frac{\lambda_3}{\mu}}\left(\nabla_z (\mathcal{D} - \mathcal{C}^{k+1}) - \frac{\mathcal{Y}_3}{\mu}\right) \tag{20}$$

where $\mathcal{S}$ denotes the soft-thresholding operator

$$\mathcal{S}_\omega(x) = sign(x) max(|x| - \omega, 0) \tag{21}$$

The above optimization process can be concluded in **Algorithm 1**. As the proposed model is convex as well as the variables can be separated into two groups, the convergence of the proposed algorithm is ensured by the ADMM framework in theory.

---

**Algorithm 1** Optimization of the Proposed Model
**Input:** Matrix $\mathcal{D} \in \mathbb{R}^{m \times n \times t}$, scalar $\mu$, $\lambda_1, \lambda_2, \lambda_3, \lambda_4$
**1:** Initialization**:**
**2: while** not converged **do**
**3:**    Update $\mathcal{A}$ via equation (10);
**4:**    Update $\mathcal{C}$ via equation (14);
**5:**    Update $\mathcal{H}$, $\mathcal{V}$ and $\mathcal{T}$ via equation (18), equation (19), and equation (20);
**6:**    Update $\mathcal{Y}$ via equation (8);
**7: end while**
**Output**: $\mathcal{B}, \mathcal{C}$

---

*2.2. Substitution of Clean Area*

After executing the above-mentioned model, the cloud and cloud shadow element as well as the clean element are acquired. The proposed model obtains the general spectral information of the images as clean element $\mathcal{B}$, while the differences between images are regarded as the cloud and cloud shadow element $\mathcal{C}$. Since the clean element is obtained from the original cloud-contaminated image and the cloud with cloud shadow has been removed, the clean element can be seen as the preliminary recovery results. Generally, the larger difference appears in the cloud area and cloud shadow area rather than the clean area. This means the cloud area and cloud shadow area can be obtained by the means of thresholding. To be specific, if the value of a pixel in the cloud and cloud shadow element is greater than the cloud threshold, this pixel will be considered as a cloud pixel, if the value of a pixel in the cloud with cloud shadow element is smaller than the cloud shadow threshold, this pixel will be considered as a cloud shadow pixel. The cloud threshold and cloud shadow threshold are given empirically. By doing this, the specific cloud and shadow area $\Omega$ and the clean area $\Omega^-$ can be acquired.

As the sparse representation-based methods process the whole image at the same time, the pixels in the original clean area can be changed and affect the subsequent processes and application. To address this problem, the clean area in the original cloud-contaminated images is used to fill the corresponding area of the preliminary results. As shown in Figure 6, to facilitate understanding, $I_{oi}$ is used to denote the original cloud-contaminated image in $i$ th time node, and $I_{Bi}$ denotes the corresponding image obtained by tensor $\mathcal{B}$. The pixels in $I_{Bi}^{\Omega^-}$ is replaced by $I_{oi}^{\Omega^-}$. The substitution step can be represented as

$$I_{Bi}^{\Omega^-} = I_{oi}^{\Omega^-} \tag{22}$$

To convenience, the procedure of tensor optimization followed by substitution is called as TOS.

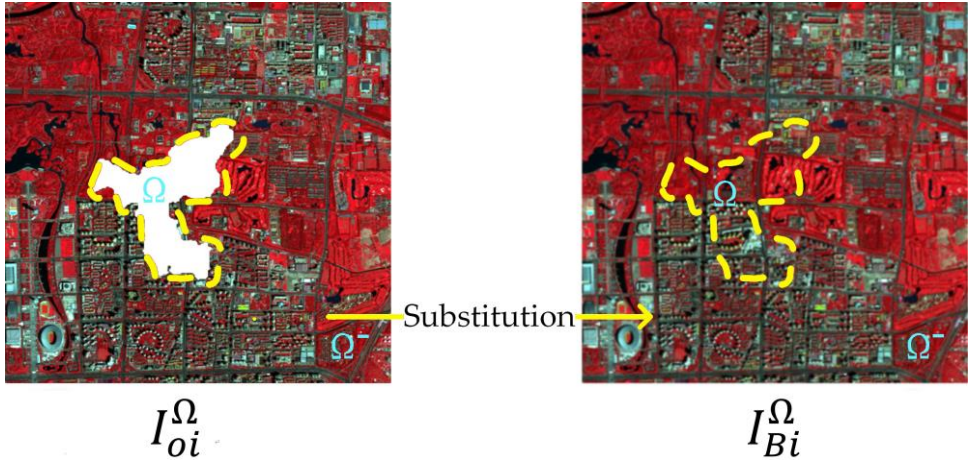

**Figure 6.** Explanation of substitution.

*2.3. Details Reconstruction*

After restoring the cloud and cloud shadow area preliminarily and replacing the original cloud-free area, the detailed information of the cloud area and cloud shadow area is reconstructed by information cloning. The image that is closest in the time dimension and that is clean in the corresponding area is taken as the reference image. Inspired by the conception of Poisson editing, the problem is modeled as a Poisson equation. To make it easier to explain, $M'$ is used to denote the reference image, and $M$ denotes the result of TOS. Let $\Omega$ denote the recovery area in the result of TOS, and the boundary is denoted as $\partial\Omega$. The intensity function to be calculated is denoted as f. $f'$ denotes the intensity function defined $M$ minus the recovery area $\Omega$. The guidance vector field defined over $M_\Omega$ is denoted as $W$. The problem can be an optimization problem with the boundary condition $f|_{\partial\Omega} = f'|_{\partial\Omega}$:

$$\min_f \iint_\Omega |\nabla f - W|^2, \text{ with } f|_{\partial\Omega} = f'|_{\partial\Omega} \tag{23}$$

where $\nabla$ represents the gradient operator, which can be calculated as follows.

$$\frac{\partial f(x,y)}{\partial x} = f(x+1,y) - f(x,y)$$
$$\frac{\partial f(x,y)}{\partial y} = f(x,y+1) - f(x,y) \tag{24}$$

Equation (23) can be solved as a Poisson equation with Dirichlet boundary conditions,

$$\Delta f = div\, W \text{ over } f, \text{ with } f|_{\partial\Omega} = f'|_{\partial\Omega} \tag{25}$$

where $div\, W$ represents the divergence of the vector field $W$, $\Delta$ represents the Laplacian operator, which defined as $\Delta = (\partial^2/\partial x^2) + (\partial^2/\partial y^2)$.

$$\frac{\partial^2 f(x,y)}{\partial x^2} = f(x+1,y) + f(x-1,y) - 2f(x,y)$$
$$\frac{\partial^2 f(x,y)}{\partial y^2} = f(x,y+1) + f(x,y-1) - 2f(x,y) \tag{26}$$

where $(x,y)$ represents the position of the pixel in the image.

The essential formulary of information cloning is shown in Equations (23) and (25), which is an optimization problem constrained by boundary conditions. The boundary condition enforces the consistency of the boundary by constraint formulation $f|_{\partial\Omega} = f'|_{\partial\Omega}$. Minimization of the $\ell^2$ norm represents that the target gradient should be as close as the guidance gradient field. By solving this problem, a seamless cloud-free and cloud shadow-free image can be obtained.

The equation (23) and (25) are discretized using the pixel grid. The 4-neighborhood of the pixel $p$ in $M$ is represented as $N_p$, and one of the pixels in 4-neighborhood is represented as $q$. $f(p)$ denotes the intensity function value at pixel $p$. Equation (23) can be rewritten as follows.

$$\min_f \sum_{<p,q>\cap\Omega\neq 0}(f(p)-f(q)-w_{pq})^2, \text{ with } f(s)-f^*(s) \text{ for all } s\in \partial\Omega \tag{27}$$

where $w_{pq}$ denotes the directional gradient of $M$ at pixel $p$.

According to (27), the following equation can be obtained:

$$|N_p|f(p)-\sum_{q\in N_p\cap\Omega}f(q)=\sum_{q\in N_p\cap\partial\Omega}f^*(q)+\sum_{q\in N}w_{pq} \text{ for all } p\in\Omega \tag{28}$$

where $|N_p|$ is the number of neighbors in $N_p$. The above equation can be solved iteratively until $f$ is converged. Because information cloning uses pixel gradient instead of pixel intensity to process the image, the recovery area looks more realistic.

## 3. Experimental Results and Analyses

In this part, seven groups of experiments that include three simulated experiments and two real-data experiments are conducted on different satellite images with different resolutions and different land features. To verify the performance of the proposed TSSTO method from various aspects, the results of TSSTO are compared with the result of the common tensor completion method, FaLRTC and HaLRTC [13], the RPCA-based method TRPCA [28], and TVLRSDC [38]. For simulated experiments, the cloud area is simulated on the original clean images. The results of the methods are compared with the original clean images from both visual effect and quantitative assessment. To evaluate the performance of TSSTO for cloud removal in simulated experiments, the following quantitative indices, peak signal-to-noise ratio (PSNR), structural similarity (SSIM) [41], cross-entropy (CE), and figure definition (FD) are employed. For real-data experiments, two datasets covered with real cloud and cloud shadow from different sensors are used to test the performance of TSSTO on real cloud-contaminated images. Since there is no original cloud-free image in the real-data experiments, only single image-based indicators can be used for analysis. Standard deviation (SD), figure definition (FD) and information entropy (IE) are calculated to quantitatively evaluate the proposed TSSTO method. Besides, cloud and cloud shadow detection results are presented to further analyze TSSTO.

*3.1. Simulated Experiments*

In this part, the cloud areas in different shapes and sizes are simulated on the original clean images that are from different sensors and with different resolutions. Experiments on three groups of simulated cloud-contaminated images are designed to evaluate the proposed TSSTO method's effectiveness and its sensitivity to cloud size. It is worth noting that PSNR, SSIM, CE, and FD are calculated using the whole image with all the bands used in the experiment.

3.1.1. Effectiveness of the Proposed TSSTO Method

In this section, experiments are performed on datasets from Gaofen-1 WFV images, and SPOT-5 images, respectively. And the results of the proposed TSSTO method are compared with those of FaLRTC, HaLRTC, TRPCA, and TVLRSDC in both qualitative and quantitative ways.

The first simulated experiment is conducted on 6 Gaofen-1 Wide Field View Multispectral (WFV) images with a size of 512×512 pixels, three bands (bands 1, 2, and 3) of which are used to undertake the test. The images mainly contain the urban area and vegetation, and the images are acquired in October 2015 and December 2015, in Hubei. The original and simulated images of true color composition (red: band 3; green: band 2; blue: band 1) are shown in Figure 7a and Figure 7b. The recovery results of FaLRTC, HaLRTC, TRPCA, TVLRSDC, TOS, and the proposed TSSTO method are presented in Figure 7c–7h. And the zoomed detailed regions of images in Figure 7 are shown in

Figure 8. The quantitative assessment indexes of the above methods are listed in Table 1, where the values of the highest accuracies are in bold.

It can be seen that the ground features are more complex and the details are richer in these images than those in the first simulated experiment. There is a huge difference between the result of TRPCA (Figure 7e) and the original image(Figure 7a), especially the city area in the left bottom. The reason is that TRPCA changes the values of the pixels in the cloud-free area. For the HaLRTC method, as shown in Figure 7d and Figure 8d, it is obvious that the texture information of the result is lost, since no information out of the image is used to reconstruct the cloud area. For FaLRTC, some details are lost in the result which does not achieve satisfactory performance, as FaLRTC utilizes no supplementary information from the reference images. The result of TVLRSDC (Figure 8f) looks more like the original image than those of the above three contrast methods(Figure 8c-e), the ground features are recovered generally and the spectral consistency is maintained nicely. In Figure 8, compared to the result of TVLRSDC, the result of TSSTO not only achieves spectral consistency with the original image but also restores more details. The texture in the yellow rectangle in Figure 8h is clearer than that in Figure 8f, and the quantitative indexes of the result of TSSTO achieve optimal among all the methods. Besides, the result of TOS restored spectral characteristics generally, which shows the reliability of the proposed model that uses UTV regularizer to constraint the temporal smoothness of the clean element.

**Table 1.** The PSNR, SSIM, AG, CE of the second simulated experimental results.

| Index | FaLRTC | HaLRTC | TRPCA | TVLRSDC | TOS | TSSTO |
|---|---|---|---|---|---|---|
| PSNR | 35.80 | 35.83 | 24.27 | 46.29 | 47.25 | **55.84** |
| SSIM | 0.9714 | 0.9716 | 0.6936 | 0.9982 | 0.9956 | **0.9998** |
| CE | $5.04 \times 10^{-4}$ | $2.80 \times 10^{-4}$ | $3.98 \times 10^{-1}$ | $5.55 \times 10^{-5}$ | $9.20 \times 10^{-5}$ | $\mathbf{3.37 \times 10^{-5}}$ |
| FD | 1204.27 | 1204.36 | 671.69 | 1213.5 | 1205.14 | **1220.96** |

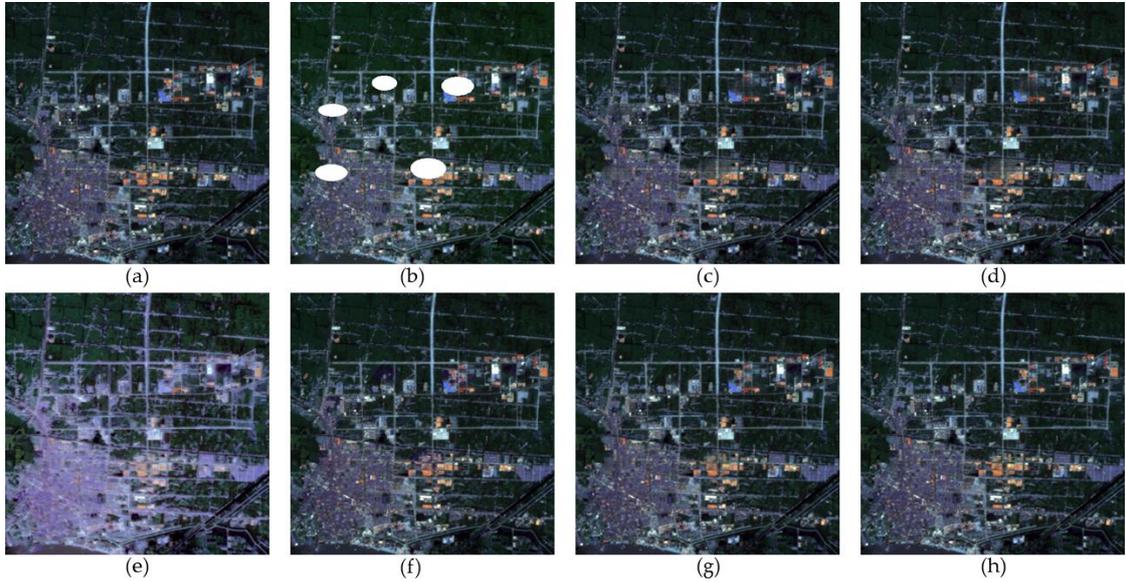

**Figure 7.** Gaofen-1 WFV images for the second simulated experiment: (a) original clean image; (b) simulated cloud image; (c)–(h) are the results of FaLRTC, HaLRTC, TRPCA, TVLRSDC, TOS, and TSSTO, respectively.

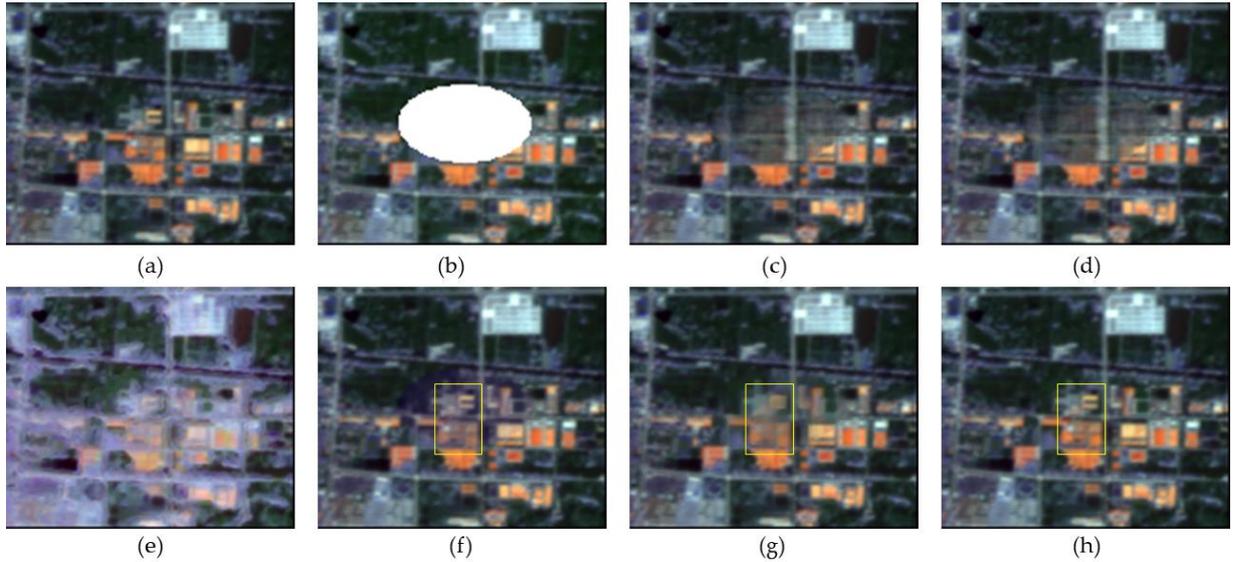

**Figure 8.** (a)–(h) are detailed regions clipped from Figure 7. (a)–(h).

In the second simulated experiment, the proposed TSSTO method is tested on 8 SPOT-5 images with four bands (bands 1, 2, 3, and 4) which are acquired in 2015, in Beijing. The size of the experimental images is 512×512 pixels. The original images of false-color composition (red: band 3; green: band 2; blue: band 1) are shown in Figure 9a. Figure 9b is the cloud-contaminated image simulated by Figure 9a. The result of the proposed TSSTO method is shown in Figure 9h. To make a comparative analysis, the proposed TSSTO method is compared with the method of FaLRTC, HaLRTC, TRPCA, TVLRSDC, and the recovery results of each contrast method are shown in Figure 9c-f. The result of TOS is also represented in Figure 9g. Figure 10a-h show the zoomed regions cropped from Figure 9a-h. The quantitative assessment indexes of the above methods are calculated within four bands and listed in Table 2 where the values of the highest accuracies are in bold.

For the results of FaLRTC (Figure 10c) and HaLRTC (Figure 10d), the reason for the blurry resulting images is that it is hard to restore a large amount of texture and structure information using little information from the cloud-free part of the cloud-contaminated image. From Table 2, the result of TRPCA gets unsatisfactory indexes, due to the differences in the non-cloud area between the original image and the resulting image (e.g. the area in the yellow rectangle in Figure 9e). In the result of TVLRSDC (Figure 10f), the recovery region looks darker than the rest of the image. This may result from that the original fourth-order dataset $\mathcal{A} \in \mathbb{R}^{m \times n \times b \times t}$ are reshaped into a matrix $A \in \mathbb{R}^{mn \times bt}$, which leads to equal treatment of information from the spectral domain and temporal domain. Besides, with matrix decomposition denoising image and removing cloud, images can be smooth and vague, but there is no processing in the TVLRSDC method to recover the details. For the result of TOS, the values of PSNR, SSIM, and FD are greater than the results of FaLRTC, HaLRTC, and TRPCA. The reconstructed result of TSSTO is more persuasive with enough details. As shown in Table 2, the quantitative assessments of TSSTO expound the validity of the proposed method as well. The result shows the effectiveness of the temporal smoothness regularized tensor optimization, which can acquire the cloud removal result that keeps spectral consistency.

**Table 2.** The PSNR, SSIM, AG, CE of the third simulated experimental results.

| Index | FaLRTC | HaLRTC | TRPCA | TVLRSDC | TOS | TSSTO |
|---|---|---|---|---|---|---|
| PSNR | 43.20 | 43.42 | 36.30 | 49.98 | 47.75 | **52.25** |
| SSIM | 0.9644 | 0.9634 | 0.9010 | 0.9976 | 0.9943 | **0.9989** |
| CE | $3.24 \times 10^{-4}$ | $1.55 \times 10^{-3}$ | $6.45 \times 10^{-2}$ | $8.59 \times 10^{-5}$ | $2.84 \times 10^{-4}$ | $\mathbf{8.09 \times 10^{-5}}$ |
| FD | 1153.77 | 1144.97 | 940.56 | 1159.46 | 1148.95 | **1171.05** |

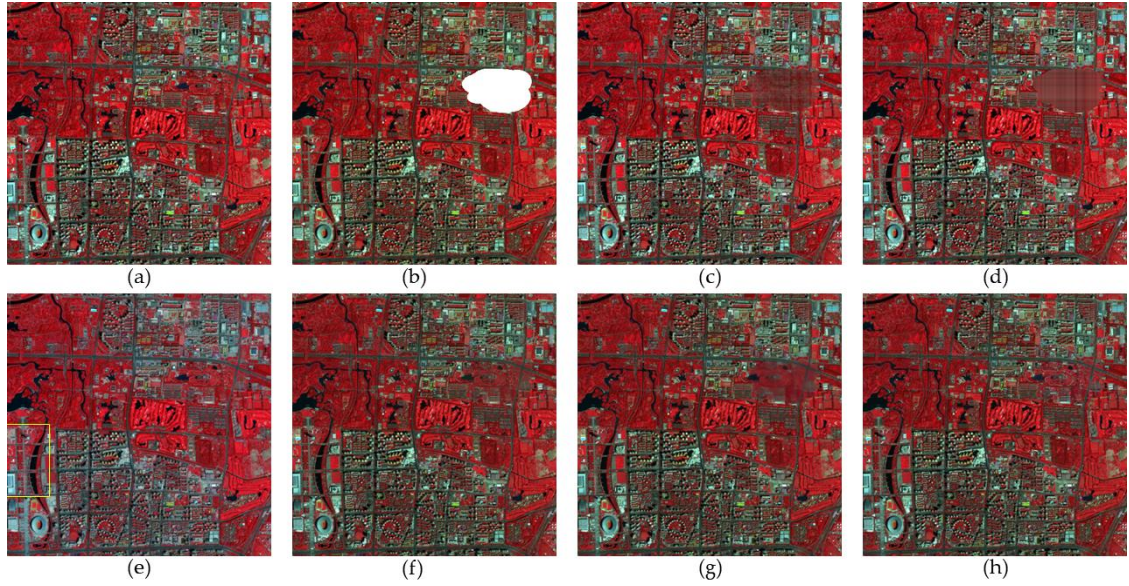

**Figure 9.** SPOT-5 images for the third simulated experiment: (a) original clean image; (b) simulated cloud image; (c)–(h) are the results of FaLRTC, HaLRTC, TRPCA, TVLRSDC, TOS, and TSSTO, respectively.

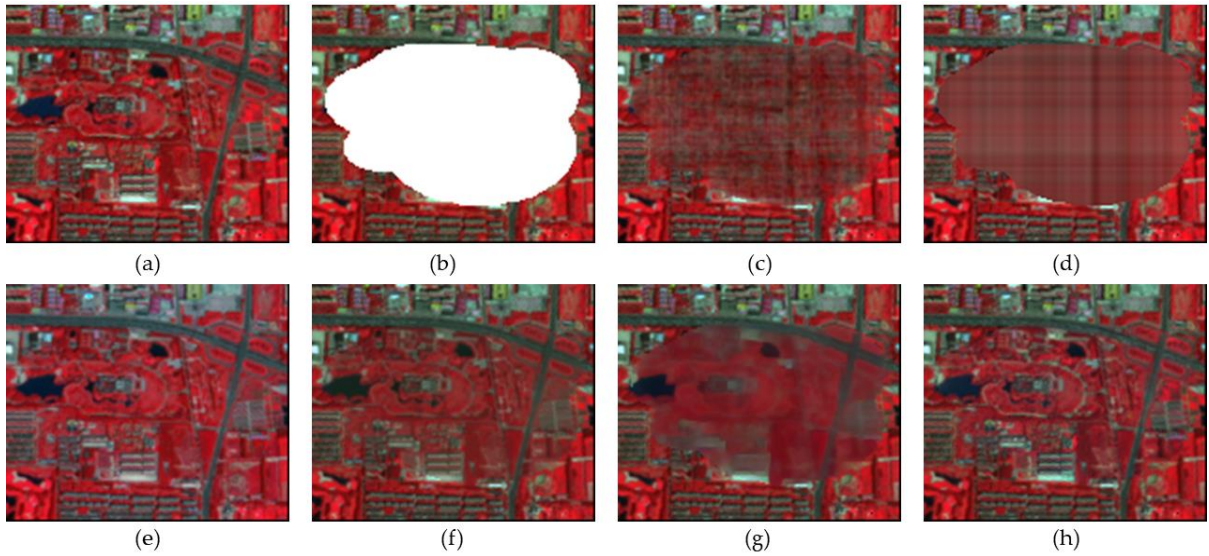

**Figure 10.** (a)–(h) are detailed regions clipped from Figure 9. (a)–(h).

3.1.2. Sensitivity to Cloud Size

In this section, the insensitivity of the proposed method to the cloud size is proved. The experiment is carried out on 11 well-aligned Landsat-8 OLI images with a size of 512× 512 pixels, three bands (bands 2, 3, and 4) of which are used to conduct the test, with different sizes of cloud area. With true color composition (red: band 4; green: band 3; blue: band 2), the simulated images and results of TSSTO are shown in Figure 11a and Figure 11b. The proportions of cloud pixels in the simulated images are 1.34% in Figure 11a, 3.92% in Figure 11b, 10.35% in Figure 11c, 19.74% in Figure 11d, 34.84% in Figure 11e, respectively. The simulated images with the results of different cloud sizes are shown in Figure 11. And the zoomed details region of Figure 11 is shown in Figure 12. Figure 13 shows the PSNR values of the results with different methods and cloud sizes.

From Figure 11-13, it can be analyzed as follows. First, since HaLRTC and FaLRTC both reconstruct the cloud-contaminated pixel using information from the cloud-free area of the target image, the curves of PSNR values of HaLRTC and FaLRTC is similar. Second, the proposed TSSTO method can achieve reasonable results of cloud removal. As shown in Figure 12e, the proposed TSSTO method can obtain detail-rich and seamless results of cloud removal, even the proportion of cloud in test image reaches about 35%. Third, despite the PSNR value declines with the increasing of the cloud sizes, PSNR values of TSSTO are always larger than those of the compared methods. Fourthly, as shown in Figure 13, it is distinct that the performance of TRPCA is poor, which can be explained as follows: HaLRTC, FaLRTC, TVLRSDC, and TSSTO change the pixel values in the cloud-contaminated area only, but TRPCA manipulates the cloud-free pixels twice and introduces errors into the cloud-free area. Through the above experiment and analyses, the insensitivity of TSSTO to cloud size is proved. With sparsity norm with two regularizers along horizontal direction and vertical direction applied on cloud and cloud shadow, TSSTO is able to obtain satisfactory results regardless of the size of the cloud area.

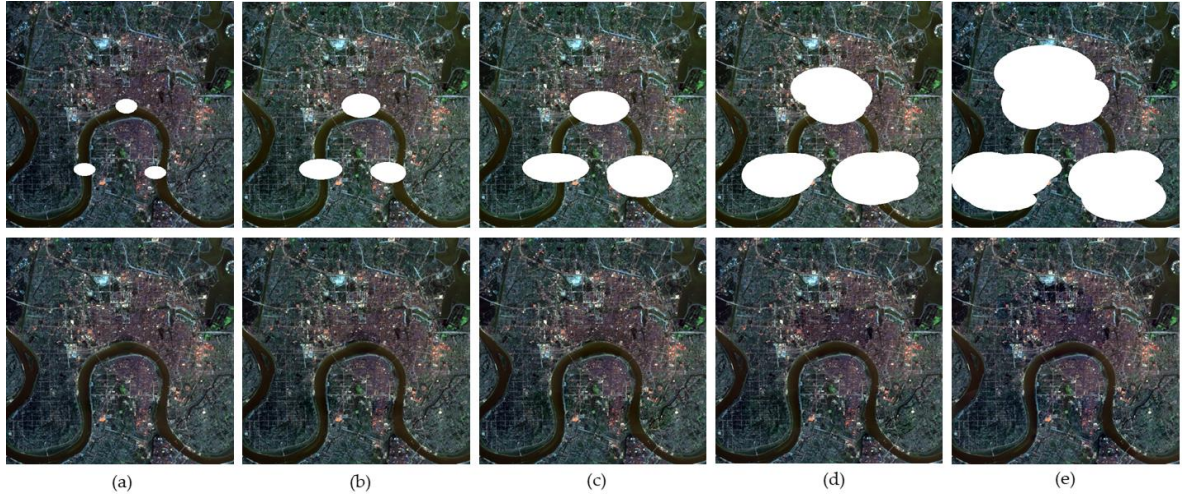

**Figure 11.** The simulated images and results with different cloud sizes. The proportions of cloud pixels in the simulated images are 1.34%, 3.92%, 10.35%, 19.74%, 34.84%, respectively.

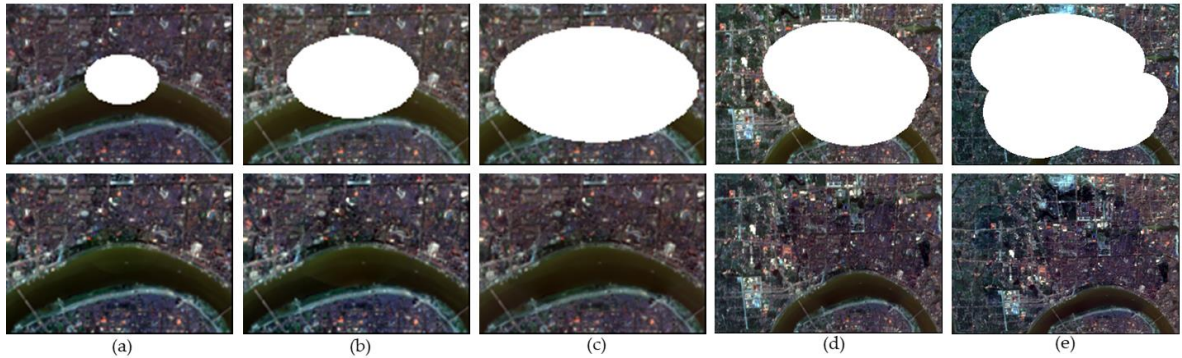

**Figure 12.** (a)–(e) are detailed regions clipped from Figure 11. (a)–(e).

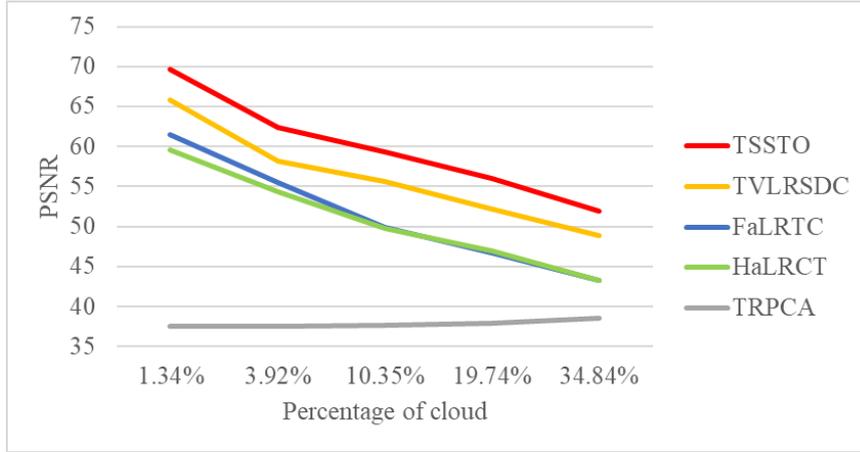

**Figure 13.** The PSNR values of the results with different methods and cloud sizes.

*3.2. Real-Data Experiments*

In this section, two groups of real-data experiments are conducted on different real cloud-contaminated images to verify the practicability of the proposed TSSTO method. The proposed method is tested on the real cloud-contaminated images from different sensors, and the cloud areas are in different shapes and different sizes. The results of TSSTO are compared with those of FaLRTC, HaLRTC, TRPCA, and TVLRDC. To evaluate the cloud removal results quantitatively, standard deviation (SD), figure definition (FD), and information entropy (IE) are calculated and summarized in Table 3. It is worth noting that these indicators are only calculated in the repaired area. To make a comprehensive evaluation of the proposed TSSTO method, the cloud and cloud shadow detection results are provided and compared with those of other methods.

3.2.1. Real-Data Experiments Results

The first experiment in this section is performed on 5 Gaofen-1 WFV images with four bands (bands 1, 2, 3, and 4) and a 1000×1000-pixel size, which are acquired in October 2015 and December 2015, in Hubei. Figure 14a shows the original image (true color composition red: band 3; green: band 2; blue: band 1). Figure 14f shows the recovery results of the proposed TSSTO method, and the results of FaLRTC, HaLRTC, TRPCA, TVLRSDC are listed in Figure 14b–14e. Figure 15 shows the zoomed detailed regions of images in Figure 14. Table 3 lists the SD values, FD values, and IE values of the results of different methods.

As shown in Figure 15, the results of HaLRTC and HaLRTC are not adequate for the demands, due to the difficulty to restore the texture in the cloud-contaminated area only using the information in the clean area. The cloud-free area (e.g. area in the yellow rectangle) in the result of TRPCA (Figure 14d) looks different from the same area in the original image (Figure 14a), due to dual RPCA applied. For the zoomed detailed result of TVLRSDC (Figure 15e), there is some apparent color deviation in the recovery area, as the information in the spectral domain and the temporal domain is processed together and the details are not specially recovered. The result of TSSTO (Figure 14f) is satisfactory visually, the zoomed area shows that TSSTO achieves optimal results with clear texture and enough details. In Table 3, compared with the results of contrast methods, the quantitative index values of the result of TSSTO is optimized. The experiment results testify the effectiveness of the proposed model that ensures temporal smoothness between images by UTV regularizer along the temporal direction, and band-by-band processing avoids large spectral differences.

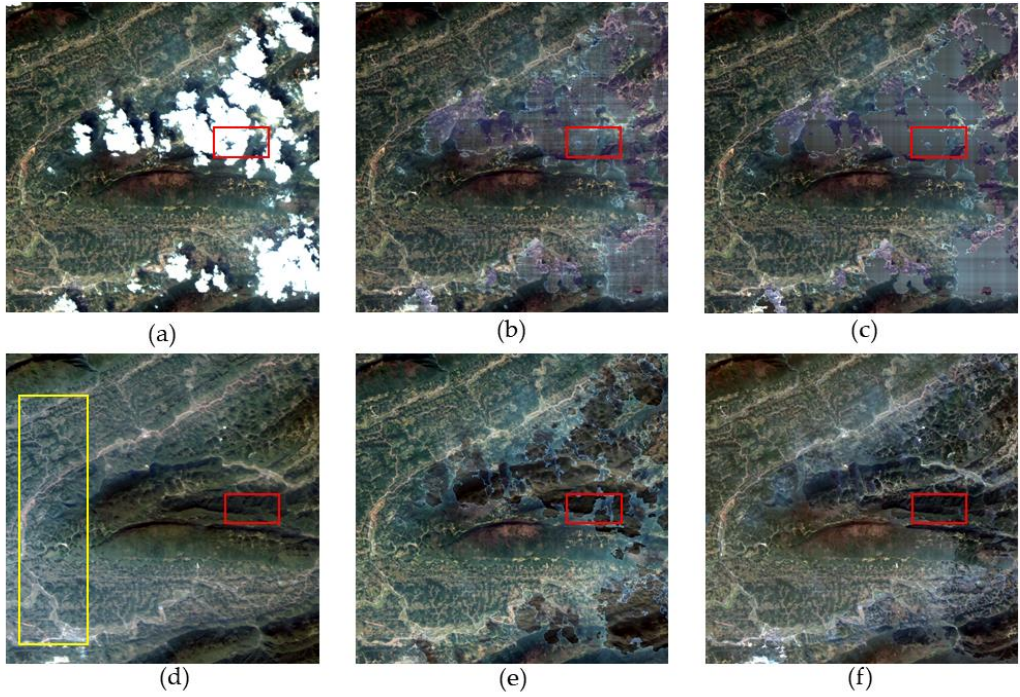

**Figure 14.** Gaofen-1 WFV images for the second real-data experiment: (a) cloud image; (b)–(f) are the results of FaLRTC, HaLRTC, TRPCA, TVLRSDC, and TSSTO, respectively.

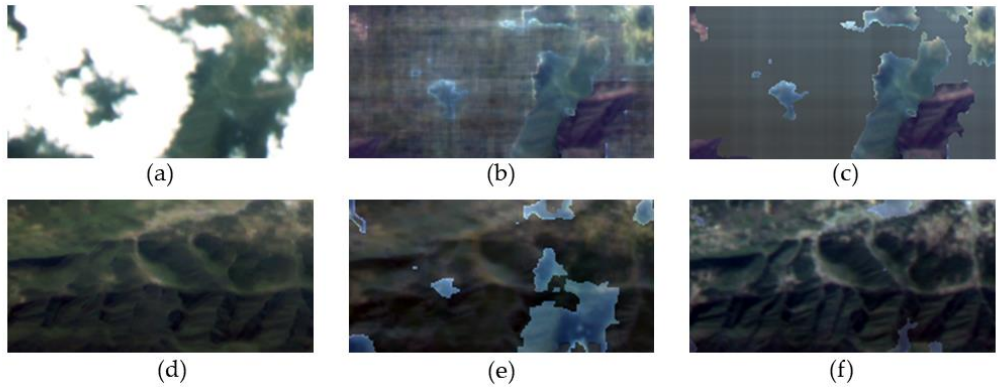

**Figure 15.** (a)–(f) are detailed regions clipped from Figure 14. (a)–(f).

In the second real-data experiment, the proposed TSSTO method is tested on 10 well-aligned SPOT-5 images with a size of 2000×2000 pixels. The images are acquired in 2015, in Beijing. Four bands (bands 1, 2, 3, and 4) of the images are used to conduct the test. The original cloud-contaminated image and results of different methods are shown in Figure 16 (red: band 3; green: band 2; blue: band 1). Figure 16a is the original cloud-contaminated image. Figure 16b-f are the results of FaLRTC, HaLRTC, TRPCA, TVLRDC, and the proposed TSSTO method, respectively. The zoomed detailed regions of images in Figure 16 are represented in Figure 17. The values of SD, FD, and IE of the results of different methods are listed in Table 3.

As shown in Figure 16-17, the results of FaLRTC and HaLRTC are not satisfactory, as large unrepaired area and complex texture information make it difficult to reconstruct the cloud-contaminated area. For the results of TRPCA, TVLRSDC, TSSTO, which are with reasonable visual effects, it is noticed that the result of TRPCA has large spectral differences from the original image (Figure 16). The transition between the original clean area and the recovery area is not smooth in the result of TVLRDC, as shown in Figure 17e. This can be explained as follows. There are large differences in the cloud area between the temporal images, while the differences in the cloud area between different bands within an image is small. Hence, treating the information from the temporal dimension and spectral dimension together can make the result unsatisfactory. Compared with the result of TVLRSDC, the result of TSSTO obtains clearer texture and rich details, as the proposed model guarantees the general spectral information to accord with the cloud-free area and information cloning guarantees plenty of details in the cloud recovery area.

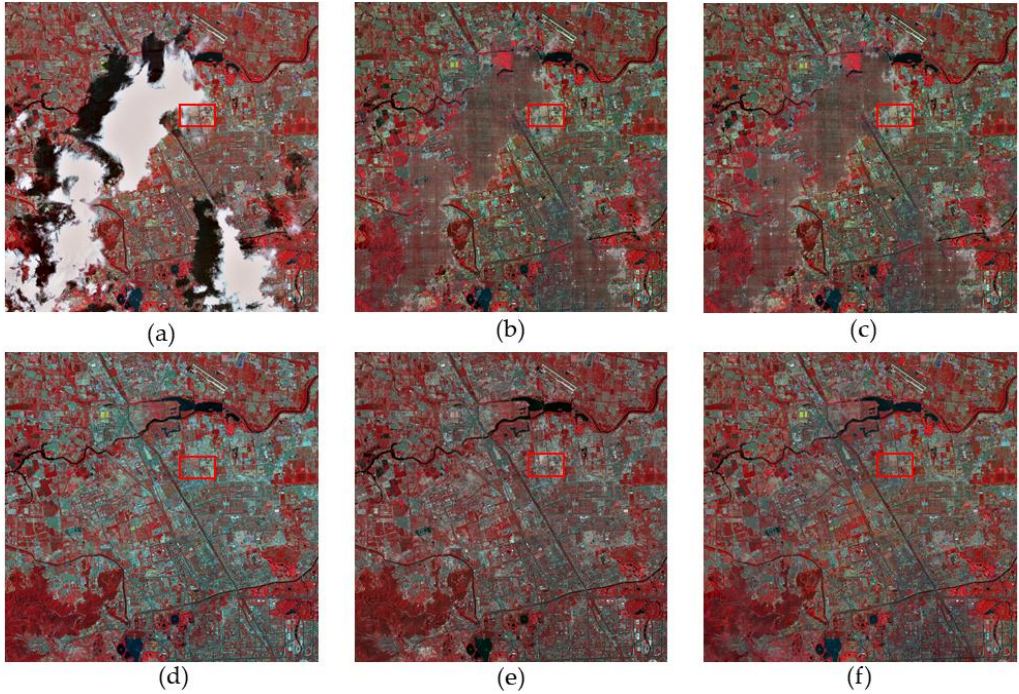

**Figure 16.** SPOT-5 images for the third real-data experiment: (a) cloud image; (b)–(f) are the results of FaLRTC, HaLRTC, TRPCA, TVLRSDC, and TSSTO, respectively.

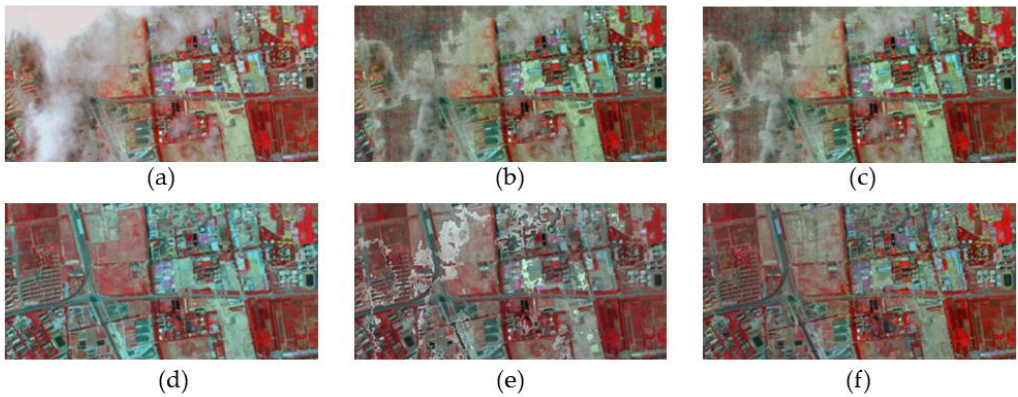

**Figure 17.** (a)–(f) are detailed regions clipped from Figure 16. (a)–(f).

**Table 3.** The values of SD and IE of the real-data experimental results.

|  |  | FaLRTC | HaLRTC | TRPCA | TVLRSDC | TSSTO |
|---|---|---|---|---|---|---|
| Figure 14 | SD | 5.260407 | 5.191628 | 5.323994 | 5.231277 | **5.726011** |
|  | FD | 1387.7772 | 1387.7772 | 1462.2249 | 1352.7260 | **1789.9368** |
|  | IE | 10.86891 | 10.86682 | 11.77331 | 10.63315 | **14.8556** |
| Figure 16 | SD | 22.3821 | 22.3829 | 8.6390 | 20.5975 | **22.8344** |
|  | FD | 2615.7688 | 2615.1677 | 2468.5109 | 2462.3352 | **2615.7837** |
|  | IE | 6.2072 | 6.2075 | 5.9784 | 6.0592 | **6.2769** |

In this chapter, to testify the performance of the proposed method, a series of comprehensive experiments are conducted on different datasets. Due to the abundant and various spatial details, the tensor complement method, HaLRTC and FaLRTC cannot reach stable results, in which the reconstructed area appears lack of texture, especially when the cloud area and cloud shadow area contain complex land cover (e.g. river and artificial facilities). TRPCA changes the values of pixels in the original clean area, which makes the result untrustworthy. TVLRSDC can process the information from spectral and temporal domains together, however, the information from different domains is treated equally without discrimination, which may cause abnormal spectral differences. For the proposed TSSTO method, with the recovery of the details in place, the result is optimal in most cases. The visual effect, as well as quantitative indexes, shows that tensor optimization can get the nearly acceptable result and information cloning restore the details effectively.

3.2.2. Cloud and Cloud Shadow Detection Result

To further evaluate the performance of the proposed TSSTO methods, the cloud and cloud shadow detection results of TSSTO with reference masks are presented in the following. For Gaofen-1 WFV satellite data, the state-of-the-art multi-feature combined cloud and cloud shadow detection method (MFC) [42] is applied to the cloud-contaminated image of Gaofen-1 WFV satellite to get reference cloud and cloud shadow mask. The cloud and cloud shadow mask from SPOT-5 image obtained by MAJA processor (http://www.cesbio.ups-tlse.fr/multitemp/?p=6203) is easily accessible from the website (https://take5.theia.cnes.fr/atdistrib/take5/client/#/home), which is used as reference mask. Figure 18 shows the cloud and cloud shadow detection from different methods.

From Figure 18, cloud and cloud shadow detection results of TSSTO look reasonable, according to the original cloud-contaminated images. For Gaofen-1 WFV image, it can be noticed that some cloud shadow is detected as cloud-free areas in the mask obtained by MFC (e.g. some shadow pixels are not detected in the yellow rectangle). In the cloud and cloud shadow detection result of the proposed TSSTO method, the cloud and cloud shadow element is acquired by the tensor optimization and this element is processed by the threshold, which avoids the confusion between clean pixels and cloud shadow pixels. For SPOT-5 image, in the result of MAJA processor, some clean area around cloud and cloud shadow is detected as cloud. A More accurate mask is obtained by TSSTO, which enables cloud removal methods not to change the pixels in cloud-free area as much as possible. This helps to preserve the spectral characteristics of cloud-free regions.

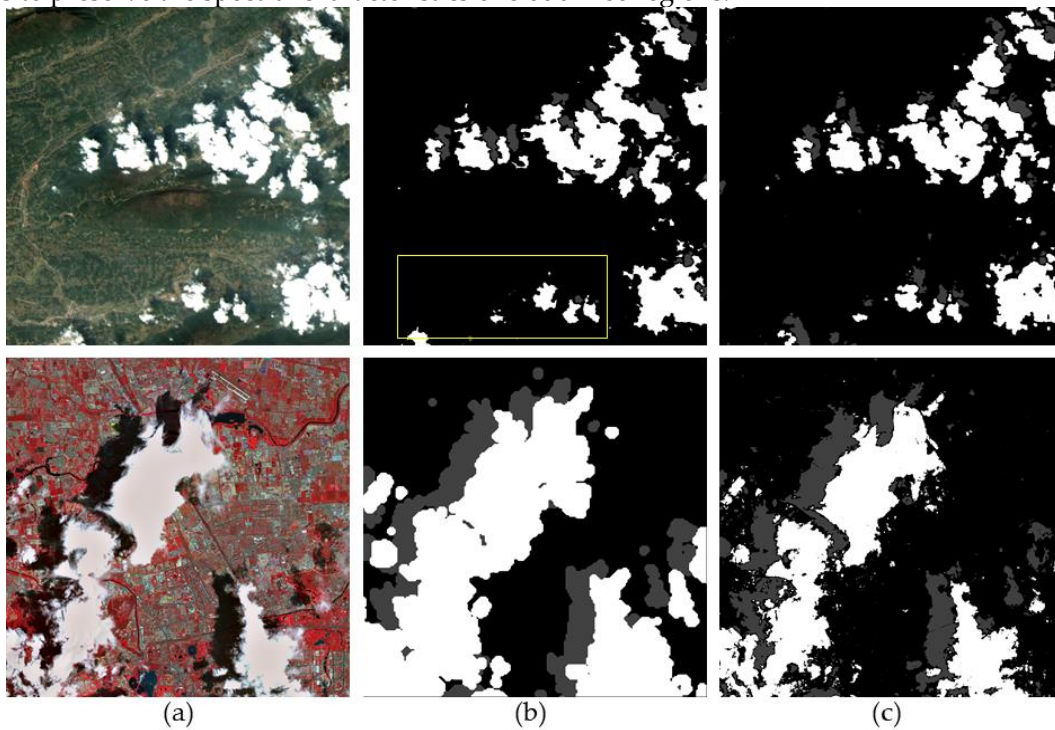

(a)          (b)          (c)

**Figure 18.** Comparison of cloud and cloud shadow masks for two real-data experiments. (a) Cloud images, (b) mask obtained by MFC method(Gaofen-1 WFV image) and MAJA processor (SPOT-5 image), and (c) masks obtained by TSSTO

## 4. Conclusions

This paper proposes a novel TSSTO method for remote sensing images cloud and could shadow removal using tensor optimization based on temporal smoothness and sparsity. The temporal smoothness of the clean element is guaranteed by the UTV regularization term in this paper which does not need to conduct SVD iteratively. For the cloud and cloud shadow element, to enhance the sparsity and ensure the smoothness along the horizontal direction and vertical direction, the cloud and cloud shadow element is constrained by the sparse norm and UTV regularizers. Besides, the convergence of the model is ensured by the ADMM framework in theory. The flow of the proposed TSSTO method encompasses tensor optimization, the substitution of cloud-free and cloud shadow-free area, and information cloning. In the tensor optimization step, the proposed model is executed, the cloud and cloud shadow element as well as the clean element is generated. In the substitution step, the cloud area and cloud shadow area are determined by the threshold segmentation. And the clean area of the original image is replaced to the clean element. Finally, information cloning is applied to restore more details in the recovery area. A series of experiments have been conducted on the different datasets, which verifies that the proposed TSSTO method not only can recover the cloud area and cloud shadow area with abundant details but also can keep spectral consistency and the continuity well. What is more, the increasing size of the cloud areas shows little influence on the proposed method.

Meanwhile, TSSTO still has room for improvement. First, the result of TSSTO is related to the result of cloud and cloud shadow detection, higher accuracy of the result of TSSTO is ensured by the credible cloud and cloud shadow mask. The authors will make a good effort to study and to overcome this limitation. Second, reliance on the quality of input images reduces some application scenarios of TSSTO. In the future, it is worth making appropriate use of the information that in spectral domain, spatial domain, and temporal domain jointly to weaken the dependency on the quality of the input images and better reconstruct the cloud-contaminated area.

**Author Contributions:** Funding acquisition, Jun Pan; Methodology, Chenxi Duan; Project administration, Jun Pan; Resources, Jun Pan; Validation, Rui Li; Writing – original draft, Chenxi Duan; Writing – review & editing, Chenxi Duan, Jun Pan and Rui Li.

**Acknowledgments:**

This work was supported by the National Natural Science Foundation of China (No. 91738301, 41971422), and land and resources research plan of Hubei Province[2018] No. 844-11

**Conflicts of Interest:** The authors declare no conflict of interest.